\documentclass[iop,apj]{emulateapj}
\usepackage{amsmath,amssymb}
\usepackage{graphicx}
\usepackage{CJK}
\usepackage{natbib}
\usepackage{color}
\usepackage[
    breaklinks,
    colorlinks=true,
    urlcolor=blue,
    linkcolor=red,
    citecolor=blue
    ]{hyperref}
\graphicspath{{./}{/Users/zjx/Documents/art/tex_et_al/}}

\usepackage[english]{babel}
\usepackage{blindtext}
\usepackage{float}

\newcommand{\chisq}{\chi^{2}}
\newcommand{\dchi}{\Delta \chi^{2}}
\newcommand{\dof}{\textrm{dof}}

\newcommand{\Ec}{E_{\textrm{c}}}

\begin{document}
\begin{CJK*}{UTF8}{gbsn}
\title{
On measuring the variation of high energy cut-off in active galactic nuclei
}
\author{
    Ji-Xian Zhang(张继贤),
    Jun-Xian Wang(王俊贤),
    Fei-Fan Zhu(朱飞凡) }
\affil{CAS Key Laboratory for Researches in Galaxies and Cosmology,
University of Science and Technology of China, Chinese Academy of Sciences, 
Hefei, Anhui 230026, China. jxzly@mail.ustc.edu.cn, jxw@ustc.edu.cn\\
School of Astronomy and Space Science, University of Science and Technology of China, Hefei 230026, China}
\email{jxzly@mail.ustc.edu.cn, jxw@ustc.edu.cn}
\begin{abstract}

The variation in the high energy cut-off E$_c$ in active galactic nuclei uniquely probes the corona physics. 
%However measuring cut-off and its variation is non-trivial as it is model dependent, and spectral fittings are often dominated by the lower energy spectral ranges where most X-ray photons are detected. 
In this work we show 
that the ratio of two NuSTAR spectra (in analogy to difference imaging technique widely used in astronomy) 
is uniquely useful in studying E$_c$ variations. The spectra ratio could directly illustrate potential E$_c$ variation between two spectra. By comparing with the ratio of two spectral fitting models, it also examines the reliability of the spectral-fitting measured E$_c$ variation.
Assisted with this technique, we revisit the 5 AGNs in literature (MCG -5-23-16, 3C 382, NGC 4593, NGC 5548 and Mrk 335) for which E$_c$ ($k$T$_e$) variations have been claimed with NuSTAR observations.
We show
the claimed E$_c$ variations appear inconsistent with the spectra ratios in three of them, thus need to be revised, demonstrating the striking usefulness of spectra ratio.
We present thereby improved spectral fitting results and E$_c$ variations.
We also report a new source with E$_c$ variations based on NuSTAR observations (radio galaxy 4C +74.26). 
We find the corona tends to be hotter when it brightens (hotter-when-brighter) in 3C 382, NGC 5548, Mrk 335 and 4C +74.27, but MCG -5-23-16 and NGC 4593 show no evidence of significant E$_c$ variations.
Meanwhile all 6 sources in this small sample appear softer-when-brighter.
Changes in corona geometry are required to explain the observed hotter-when-brighter trends.

\end{abstract}

\section{Introduction}\label{sect:intro}

It is widely believed that the hard X-ray emission in active galactic nuclei (AGN) originates in a hot and compact region, the so called corona
\citep{Haardt1991,Haardt1994,Haardt1997b}.  
The X-ray emission of AGNs usually is highly variable, on time scales as short as a few hours.
In line with X-ray reverberation mapping studies \citep[e.g.][]{Grier2012a, Peterson2004d} and micro lensing observations \citep[e.g.][]{Chartas2002a, Chartas2009b}, it indicates a small physical size of the corona, typically 
several $r_g$. 
In the standard disc-corona paradigm,  
the X-ray emission is due to thermal Comptonization of the soft disc photons by the hot coronal electrons. The process produces a power-law shaped X-ray spectrum with a high-energy cut-off. Such cut-off, a characteristic feature of thermal Comptonization, has been observed at $\approx$100 keV in a number of sources \citep[e.g.][]{Zdziarski2000,Ricci2011,Molina2013d,Brenneman2014,Malizia2014,Marinucci2014,Fabian2015,Ursini2015,Keek2016,Lubinski2016,Ursini2016,Zoghbi2017}

X-rays spectral variation provide unique clues to the physical nature of the corona, which currently remains poorly known.
Recently, changes in high energy cut-off have been detected in individual AGNs \citep[e.g.][]{Ballantyne2014, Madsen2015, Ursini2015,Keek2016,Ursini2016,Zoghbi2017}. 
The changes generally reflect variation in corona temperature.
Such variation, combined with the coordinated changes in flux and spectra slope, opens a new window to 
constrain the physical models of the corona.
 
However, measuring the corona temperature and its changes is not straightforward even using X-ray spectra obtained with the state-of-art hard X-ray imaging telescope NuSTAR \citep{Harrison2013}.  
Fitting X-ray spectra with thermal corona Comptonization models is clearly model dependent,  as the coronal physics and geometry are still poorly known.
A more general approach is to fit the spectra by an exponential cut-off power-law.
As the high energy curvature of a Comptonized spectrum could be sharper than a simple exponential cut-off \citep{Zdziarski2003}, fitting with cut-off power law thus only provides approximation and phenomenological description to the spectra.
For optically thick thermal corona, $k$T$_e$ $\approx$ E$_c/3$ ($\tau$ $\gg$ 1), and $k$T$_e$ $\approx$ E$_c/2$ for $\tau$ $\lesssim$ 1  \citep{Petrucci2001a}.
Note for a non-static corona, i.e., with relativistic bulk motion or outflow \citep{Liu2014}, the relation between E$_c$ and $k$T$_e$ could be more complicated. 

Additionally, spectral fitting could be affected by other components in the spectra which also produce continuum curvatures (such as due to complicated absorption and reflection component).
This is particularly important considering the S/N ratio above several tens keV in the spectra is very low even for the local X-ray brightest Seyfert galaxies.
As the best-fitting is dominantly determined by the spectra at lower energies where much more photons are detected, the measurement of $E_c$ or $k$T$_e$
might be significantly biased.

In this work, we show that using 
the ratios of the spectra obtained at different epochs, one can easily and model-independently
select sources with possible variations in high energy cut-off. The spectra ratios can further assist followup detailed spectral fitting to confirm and quantify the $E_c$ variations, 
as such ratios can to be used to examine the goodness of spectral fitting, in additional to the commonly adopted data-to-model residuals for individual spectrum.

In \S \ref{sect:data} we present NuSTAR observations of sources analyzed in this work and our data reduction. In \S \ref{method} we introduce the approaches to derive spectra ratios, and the validity of using such spectral ratios
to illustrate potential E$_c$ variations.  
 In \S \ref{sect:others}, we collect in literature five individual AGNs for which changes in E$_c$ (or $k$T$_e$) have been reported with NuSTAR observations.
With the help of this new technique, we revisit their E$_c$ variations and report the detection of E$_c$ variation in a new source, 4C +74.26 at $z=0.1040$ \citep{Riley1989}.
Discussion and conclusions are given in \S \ref{sect:discussion}.

\section{Observations and Data Reduction}\label{sect:data}
	\begin{table}[H]
		\begin{center}
			\caption{The logs of the NuSTAR observations analyzed in this work. \label{obs}}
			\begin{tabular}{c c c c c } 
				\hline \rule{0pt}{2.5ex} Source & Obs. & Obs. Id. & Start time (UTC)  & Net exp.\\ & & & yyyy-mm-dd & (ks)  \\ \hline \rule{0pt}{2.5ex}
				MCG 5-23-16 & A & 60001046002 & 2013-06-03 & 160 \\
				& B & 60001046004 & 2015-02-15 & 210 \\
				& C & 60001046006 & 2015-02-21 & 98 \\
				& D & 60001046008 & 2015-03-13 & 220 \\ \hline \rule{0pt}{2.5ex}
				3C 382 & A & 60061286002 & 2012-09-18 & 16 \\
				& B & 60001084002 & 2013-12-18 & 82 \\\hline \rule{0pt}{2.5ex}
				NGC 4593 & A & 60001149002 & 2013-06-13 & 23 \\
				& B & 60001149004 & 2013-06-13 & 21 \\
				& C & 60001149006 & 2013-06-25 & 21 \\
				& D & 60001149008 & 2014-09-20 & 23 \\ 
				& E & 60001149008 & 2014-09-20 & 21 \\\hline \rule{0pt}{2.5ex}
				NGC 5548 & A & 60002044002 & 2013-07-11 & 24 \\
				& B & 60002044003 & 2013-07-12 & 27 \\
				& C & 60002044005 & 2013-07-23 & 49 \\
				& D & 60002044006 & 2013-09-10 & 51 \\ 
				& E & 60002044008 & 2013-12-20 & 50 \\\hline \rule{0pt}{2.5ex}
				Mrk 335 & A & 60001041002 & 2013-06-13 & 21 \\
				& B & 60001041003 & 2013-06-13 & 21 \\
				& C & 60001041005 & 2013-06-25 & 93 \\
				& D & 80001020002 & 2014-09-20 & 68 \\ \hline \rule{0pt}{2.5ex}
				4C +74.26 & A & 60001080002 & 2014-09-21 & 19 \\
				& B & 60001080004 & 2014-09-22 & 56 \\
				& C & 60001080006 & 2014-10-30 & 90 \\
				& D & 60001080008 & 2014-12-22 & 42 \\ \hline \rule{0pt}{2.5ex}			
			\end{tabular}
		\end{center}
	\end{table}
	
NuSTAR \citep{Harrison2013} provides unpreacedented sensitivity and high spectral resolution at energies above 10 keV, ideally suited to constrain the high-energy cut-off 
in AGNs. 
NuSTAR observation logs of six AGNs studied in this work are presented in Table~\ref{obs}.

The NuSTAR data were reduced using the standard pipeline (\textsc{nupipeline}) with the NuSTAR Data Analysis Software (\textsc{nustardas}, v1.6.0; part of the \textsc{heasoft} distribution as of version 6.19), and calibration files NuSTAR {\sc caldb} v20161021. Spectra were extracted using the standard tool {\sc nuproducts} for each of the two hard X-ray detectors aboard NuSTAR, which sit inside the corresponding focal plane modules A and B (FPMA and FPMB). The source spectra were extracted from circular regions with a radius of 75\arcsec, and background from a circular ring with inner radius of 90\arcsec\ and outer radius of 130\arcsec. 
The spectra from FPMA and FPMB were analyzed jointly but not combined.
The spectra were grouped to have a minimum of 50 counts per bin.

Spectral fitting to $3-78$ keV was carried out with the {\sc xspec} 12.9 package \citep{Arnaud1996} and the $\chisq$ statistic. 
All errors are quoted at the 1$\sigma$ confidence level ($\dchi = 1.0$) for one interesting parameter, unless otherwise stated.
For some physical parameters which are consistent with non-detections within 90\% confidence level (such as absorption column density N$_H$, high energy cutoff E$_c$), 
corresponding lower or upper limits are given. 

\section{Spectra ratios reveal spectral variations}\label{method}

In astronomical X-ray spectral analyses, the observed X-ray spectrum is related to incident spectrum f(E) and  the instrument response function R(I,E), 
C(I) =$\int$ f(E) $*$ R(I,E) dE,
where R(I,E) describes the probability that an incoming photon of energy E will be detected in channel I.
At given E,  R(I,E) is a rather complicated function of channel I with broad distribution.
One generally need to fit (using certain statistics) the models folded through response function to the observed data,
and seek for models/parameters which can best describe the observations. 

However, obtaining the best-fit models and/or parameters are nontrivial, particularly for complex spectra and models.  
In additional to the statistics, the residuals of the fitting (such as data-to-model ratio plots) are useful to visually examine the goodness of the fitting and whether there are spectral features which 
may have not been well fitted. In many cases, one has to carefully re-bin the data-to-model plots to look for potential weak deviations from the best-fit models, and such weak deviations could have been missed 
in studies (see \S\ref{sect:others} for examples).

To study the spectral variations, a common approach is to fit the spectra obtained at different epochs with physical or phenomenological models, 
and study the variations of the parameters. 
It is however not easy to intuitively illustrate the spectra variations with the numerical parameters only.
Furthermore, if the spectra fitting is not sufficiently good (see \S\ref{sect:others} for examples), comparison of the models/parameters could be problematic. 

In astronomical studies, imaging subtraction methods (difference imaging, for PSF matched images) are widely used for transient event selection or variability monitoring. 
Similarly, X-ray spectra ratios could be useful to demonstrate potential spectral variations. 
Considering the instrument responses, it is not straightforward to derive the X-ray spectra ratios. 
Below we present three different approaches and look for the best for the purpose of this study. 

The simplest approach is to divide two observed spectra (using the nominal energy boundaries) after correcting the difference in instrument effective area ARF(E) between two epochs.
\begin{equation}
S_{ratio} = \frac{D_1}{D_2} \times \frac{ARF_2}{ARF_1}
\end{equation}

The second approach corrects the effects of the response matrix R(E,I) (the combination of RMF and ARF)
\begin{equation}
S_{ratio} = \frac{D_1*\int m(E)*R_2(E,I)\,dE}{D_2*\int m(E)*R_1(E,I)\,dE}
\end{equation}
where m(E) is a single powerlaw with photon index of 2.0, corresponding to  a model of constant in E$*$f(E).

The third approach allows different models for different spectra
\begin{equation}
S_{ratio} = \frac{D_1*\int m_2(E)*R_2(E,I)\,dE}{D_2*\int m_1(E)*R_1(E,I)\,dE} * \frac{m_1}{m_2}
\end{equation}

We note the third approach actually give the ratios of two unfolded spectra, as
an unfolded spectrum is calculated as
\begin{equation}
\frac{D}{\int m(E)*R(E,I)\,dE}*m(E)
\end{equation}

It is clearly known that X-ray unfolded spectra are not model-independent, and unfolded spectra can only be used with great caution.
Note the second approach actually also gives the ratio of two unfolded spectra but derived using a common constant model. Sometimes such unfolded spectrum
is considered as ``model-independent"\footnote{http://cxc.harvard.edu/sherpa/faq/no\_model.html for Sherpa and http://space.mit.edu/cxc/isis/manual.html for ISIS unfolded spectrum} though it still relies on the usage of a constant model.

Two concerns need to be addressed before we can apply these approaches on observed spectra. The first is which approach is the best for the purpose of this study, i.e., to reveal potential 
broadband spectral variations including changing of powerlaw slopes and high energy cutoffs.
The second is whether the model dependence (of the 3rd approach) could affect our analyses.

\subsection{Spectra ratios of simulated spectra}

\begin{figure}
\centering
\includegraphics[width=0.5\textwidth]{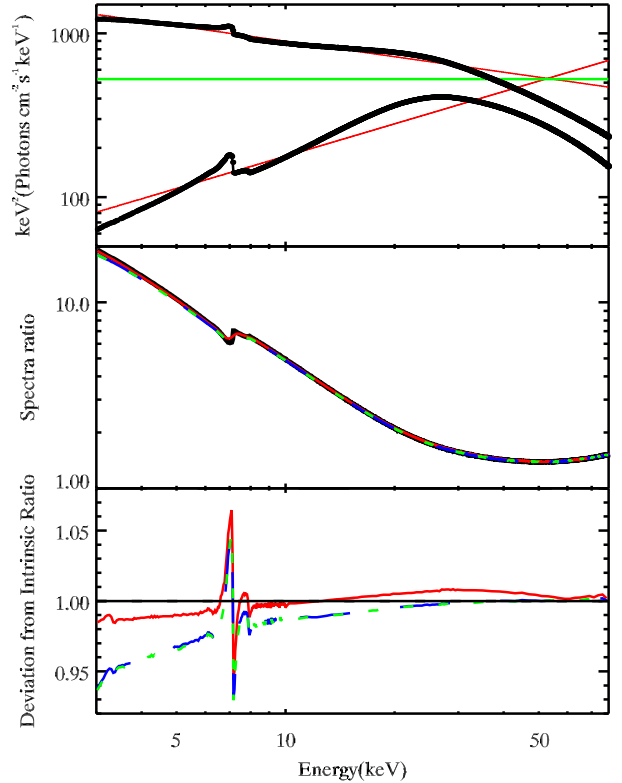}
\caption{Upper panel: two simulated NuSTAR/FPMA spectra using {\sc xspec fakeit} {\sc cutoffpl + relxill}, $\Gamma$ = 1.5 and 2.5 respectively, see text for details).  The green line is the constant model used in the 2nd approach to derive the spectra ratio. The two red lines are simple powerlaw fitting to the simulated spectra, and they are used to derive the spectra ratio following the 3rd approach. Middle panel: the intrinsic ratio of the two simulated spectra (black), and the spectra ratios recovered from three approaches (1st: blue, 2nd: green and 3rd: red respectively). Lower panel: deviation from the intrinsic ratio. 
}
\label{simspec1}
\end{figure}

In principle, an ideal approach should be simple, and able to recover the intrinsic spectral variations as accurate as possible.
The latter can be testified with simulated spectra for which the intrinsic incident spectra are known.

We utilize the NuSTAR response files of observation ID 60001046002 and 60001046004 (on MCG -5-23-16) to simulate spectra. Note selections of response files from different observations would not alter the results in this work.
Unrealistic exposure time of 10 Ms was use to simulated spectra with high S/N through out the spectral ranges. 
We first simulate spectra with two models:
1) a cutoff powerlaw with $\Gamma$=1.5, $E_c$ of 50 keV, and a relativistic disk reflection component (spec model {\sc relxill},  \citealt{Garcia2013}, for a non-spinning black, r$_{in}$ = 6 $R_g$, r$_{out}$ = 400 $R_g$, an emissivity profile index of -3.0) with a reflection fraction of 3.0; and 2) a brighter (with 50x larger powerlaw normalization\footnote{The powerlaw normalization is a multiplicative constant independent to energy. Using different values does not change the slopes of either the spectra or the spectra ratio, thus will not affect the analyses in this work.} at 1 keV, representing a softer-when-brighter pattern commonly seen in AGNs) cutoff powerlaw with $\Gamma$=2.5, $E_c$ of 150 keV, and disc reflection component {\sc relxill} with reflection fraction of 1.0 (see Fig. \ref{simspec1}).
In the upper panel of Fig. \ref{simspec1} we over-plot the models we adopted to derive the spectra ratios. For the 2nd approach, the model is a powerlaw with photon index $\Gamma$=2.0, corresponding to a constant in E$*$f(E) (green solid line). 
For the 3rd approach, 
we simply use a single powerlaw to fit each simulated spectrum, and the best-fit powerlaw photon index $\Gamma$ are 1.34 and 2.29 respectively (red solid lines). 

\begin{figure}
\centering
\includegraphics[width=0.5\textwidth]{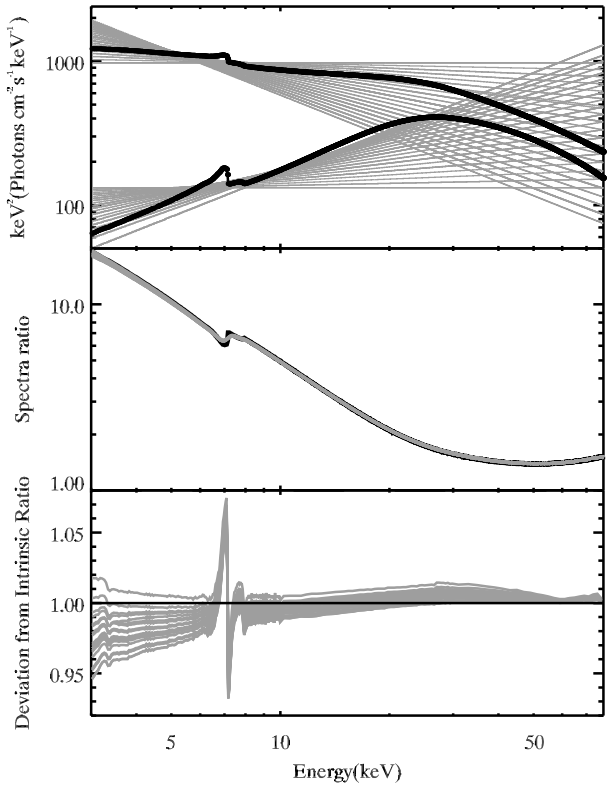}
\caption{Similar to Fig. \ref{simspec1}, but the photon index $\Gamma$ of the simple powerlaw were set at random values between 1.0 - 2.0 and 2.0 - 3.0 respectively. 
The spectra ratios derived with such powerlaw models (following the 3rd approach) show only weak scatter (middle and lower panel). 
}
\label{fake-gamma}
\end{figure}

In the middle panel we plot the spectra ratios derived through the three approaches (blue, green and red lines respectively), all of which appear consistent with the intrinsic one (in black, the ratio of the two spectral models we used to generate artificial spectra). The spectra ratios clearly demonstrate the spectral variations we input, i.e., the softening of the powerlaw (the ratio generally drops with increasing energy), and the variation of the cutoff energy (an up-tail at E $>$ 30 keV).

Weak deviations of the 3 spectra ratios from the intrinsic one (spectra ratio$/$intrinsic ratio)  are visible and illustrated in the lower panel. 
It is remarkable that the deviations are rather weak for all 3 approaches, while the 3rd approach shows the weakest deviation ($<$ 2\%, except for within the Fe K line range).
We stress that the focus of this work is the variation of the broadband spectral features including the spectral slope and high energy cutoff, and the Fe K line region could be excluded from the spectra ratio plot (note but not from followup spectra fitting). 
In Fig. \ref{simspec1}, the fact that the 3rd approach better recovers the intrinsic spectra ratio can be interpreted as following: although the two powerlaws provide poor fit to the simulated spectra (the upper panel of Fig. \ref{simspec1}), they still better describe the spectra comparing with the constant model.

\begin{figure}
\centering
\includegraphics[width=0.5\textwidth]{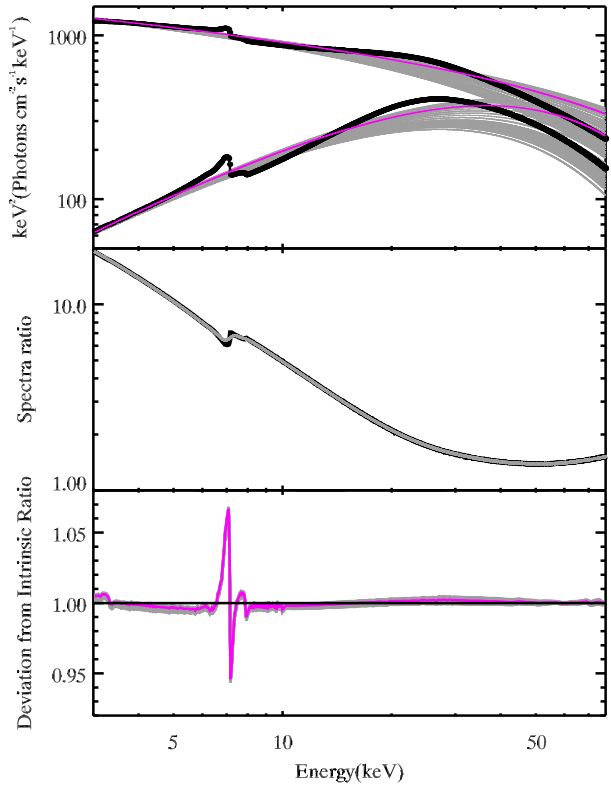}
\caption{Similar to Fig. \ref{simspec1}, but a single {\sc cutoffpl} (magenta) is used to fit the simulated spectra. Changing E$_c$ to random values (grey lines in the upper panel) yields little scatter in the derived spectra ratios
(in the lower panel). 
}
\label{bettermodel1}
\end{figure}

\begin{figure}
\centering
\includegraphics[width=0.5\textwidth]{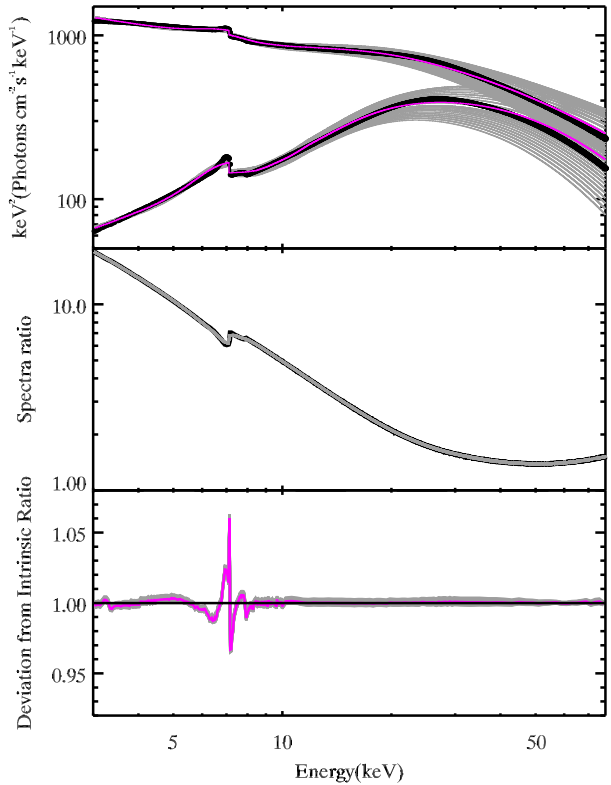}
\caption{Similar to Fig. \ref{bettermodel1}, but {\sc cutoffpl} $+$  {\sc pexrav} $+$ {\sc gauss} (magenta) is adopted to fit the simulated spectra. 
}
\label{bettermodel2}
\end{figure}

In Fig. \ref{fake-gamma}, we present powerlaw models with photon index $\Gamma$ different from the best-fit values (random values between 1.0 - 2.0 for the first spectrum, and 2.0 - 3.0 for the second one).
Spectra ratios derived with these powerlaw models following the 3rd approach show clear but rather weak scatter (grey lines in the middle and lower panel), indicating the spectra ratio derived through the 3rd approach is in fact insensitive to the model adopted. Note the scatter is even weaker  at higher energies than at lower energies.
This is  because the spectral redistribution function of X-ray detectors (the response in each pulse-height channel of photons of any particular energy) exhibits a low energy tail and low energy peaks \citep[e.g.][]{Bautz99},
thus a small but non-negligible fraction of the high energy input photons would generate low energy signals, but not vice versa.
Thus the unfolded spectrum at lower energies is more sensitive to the models adopted to derive them, comparing with that at higher energies.

We further try two phenomenological models which can further better describing the simulated spectra than the single powerlaw (cutoff powerlaw: Fig. \ref{bettermodel1}, cutoff powerlaw $+$ pexrav $+$ gauss: Fig. \ref{bettermodel2}). Here {\sc PEXRAV} models the X-ray  continuum reflection component from neutral material \citep{Magdziarz1995}.
Spectra ratios derived based on the two best-fit models both agree with the intrinsic one with deviation $<$ 1\% (again excluding the Fe K line region).
This demonstrate that spectra ratios based on models which can better describe the broadband spectra  can better recover the intrinsic spectra ratio. 
For real spectra, since there are few photons at high energy, the measurements of E$_c$ could be highly uncertain. In Fig. \ref{bettermodel1} and \ref{bettermodel2} we also show models with E$_c$ values 
different from the best-fit ones (22 - 35 keV vs. best-fit value 32 keV in Fig. \ref{bettermodel1};  and 65 -- 150 keV vs. 126 keV in Fig. \ref{bettermodel1}). The derive spectra ratios based on such models show rather weak scatter ($<$ 1\%), again demonstrating the spectra ratios derived through the 3rd approach is insensitive to the models we adopted. 
Particularly, we emphasize that prior accurate measurements of E$_c$ are not needed to derive the spectra ratio. 

\begin{figure}
\centering
\includegraphics[width=0.5\textwidth]{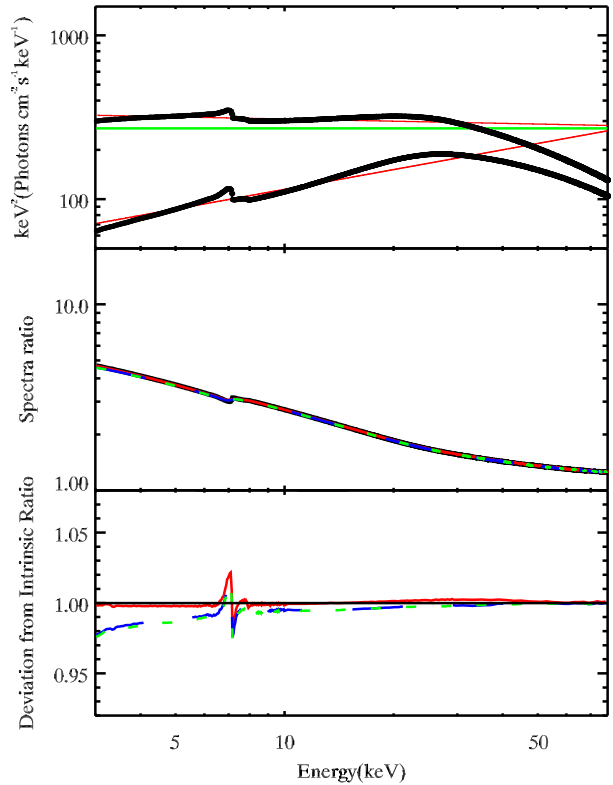}
\caption{Similar to Fig. \ref{simspec1}, but using two simulated spectra with less dramatic spectral variation (see text for details).
}
\label{simspec2}
\end{figure}

In previous analyses, we simulate two spectra with rather dramatic spectral variation ($\Gamma$ changes from 1.5 to 2.5). Such strong spectral variation is rare in reality. 
We then simulate two spectra with more realistic spectra variations: 
1) a cutoff powerlaw with $\Gamma$=1.8, $E_c$ of 100 keV, plus {\sc relxill} with a reflection fraction of 1.5; and 2) a brighter (with 5x larger powerlaw normalization) cutoff powerlaw with $\Gamma$=2.2, $E_c$ of 120 keV,  plus {\sc relxill} with reflection fraction of 1.0 (Fig. \ref{simspec2}).
Comparing with those presented in Fig. \ref{simspec1}, the deviations of the 3 derived spectra ratios from the intrinsic one are all weaker. For the 3rd approach, the deviation is $<$ 0.3\% (excluding the Fe K line region).

\subsection{Spectra ratio of MCG -5-23-16 as an example}

\begin{figure}
\centering
\includegraphics[width=0.5\textwidth]{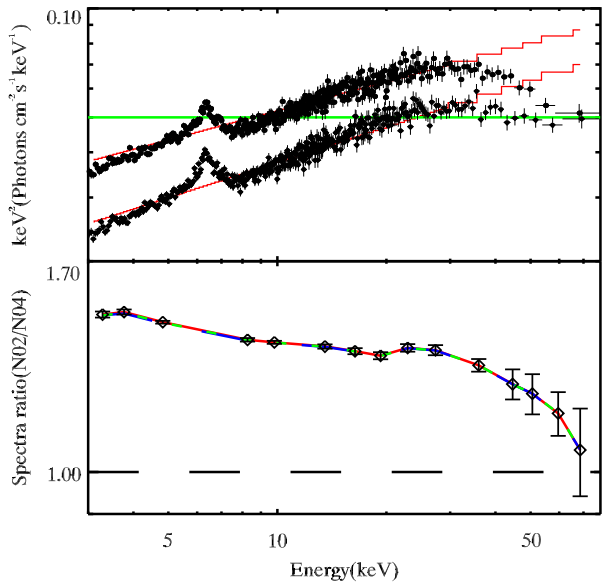}
\caption{Two NuSTAR FPMA spectra of MCG -5-23-16 (ObsID 60001046002 and 60001046004). The constant model (in E$*$f(E)) and simple powerlaws used to generate the spectra ratios are over-plotted in the upper panel. 
In the lower panel,  we plot the spectra ratio of two observations derived following the 3 approaches (data from FPMA and FPMB are combined and Fe K line region excluded, hereafter the same), and they appear consistent with each other (blue line: the 1st approach; green line: the 2nd approach;  data points and red line: the 3rd approach). Note the blue, green and red lines almost totally overlap. 
}
\label{realspec}
\end{figure}

We then plot two real NuSTAR spectra of MCG -5-23-16 in Fig. \ref{realspec}. 
Since the real spectra have much lower photon counts comparing with the simulated ones, we have to re-bin each spectrum (with identical group setting) to derive the spectra ratio. 
The re-bin strategy is 0.2 keV bin in 3-20 keV band, 0.5 keV bin in 20-50 keV band, and more than 20 net photons (source - background) each bin at $>$50 keV.
The spectra ratio we obtained is further re-binned for display purpose only. 
We stress the spectra re-binned this way were only used to illustrate the spectra ratio, and were not used for spectral fitting to derive any physical parameters. 
We also stress that selecting different re-bin strategies will not alter any conclusions presented in this work.  
The spectral variation between the two observations is moderate, and we can barely see difference in the spectra ratios derived following the 3 approaches.

We conclude that all the three approaches we discussed above can well recover the intrinsic spectra ratio, and  the 3rd approach (equation 3, with a single powerlaw model) 
has the highest accuracy ($<$ 0.3 -- 1\%, excluding the Fe K line region) among them.
 We stress that the spectra ratios derived with the 3rd approach in this work are rather insensitive to the models adopted. 
Using models more complicated than a single powerlaw which could better describe the broadband spectra would in principle yield better results, however, it is practically not needed in this work considering the photon statistics of real spectra.\footnote{It maybe be required for spectra with much higher S/N obtained with future observatories.}

\subsection{The usefulness of spectra ratios}

The spectra ratio of two spectra can 
intuitively illustrate the spectral variation without detailed modeling of each spectrum (as we have shown with simulations), in analogy to the difference imaging technique.
Furthermore, while plotting the ratio of two corresponding best-fit models could also demonstrate spectral variations,
the spectra ratio is more straightforward as it directly presents the data.  
It is particularly useful while analyzing a large sample of sources, for which accurate spectral fitting to each of them could be challenging. One can easily select candidates with possible Ec variations and perform followup detailed spectral fitting to confirm and quantify the spectral variations. 

In X-ray studies, astronomers generally examine the spectral residuals (e.g. data-to-model ratio plot) for the goodness of the fitting and possible additional spectral features.  
Similarly, for the study of spectral variation, one can compare the ratio of two spectra with that of the two corresponding best-fit models to further check the goodness of the fitting. 
Any $E_c$ variation pattern derived through spectral fitting, if inconsistent with the spectra ratio, is not likely reliable.

\section{Test on AGNs with claimed E$_c$ variations}\label{sect:others}

In literature, using NuSTAR observations (often together with simultaneous exposures from other instruments),  cut-off energy (corona temperature) variations have been claimed  in five individual AGNs, including  MCG -5-23-16 \citep{Zoghbi2017}; 3C 382 \citep{Ballantyne2014}; NGC 4593 \citep{Ursini2016};  NGC 5548 \citep{Ursini2015} and Mrk 335 \citep{Keek2016}.

Assisted with the spectra ratio technique, a systematic study of us on E$_c$ variations in a large sample of AGNs is underway. In this work we first apply this technique on
these AGNs with claimed E$_c$ variations, to demonstrate its usefulness and revisit the published E$_c$ variation patterns.

\begin{figure*}
\centering
\includegraphics[width=0.9\textwidth]{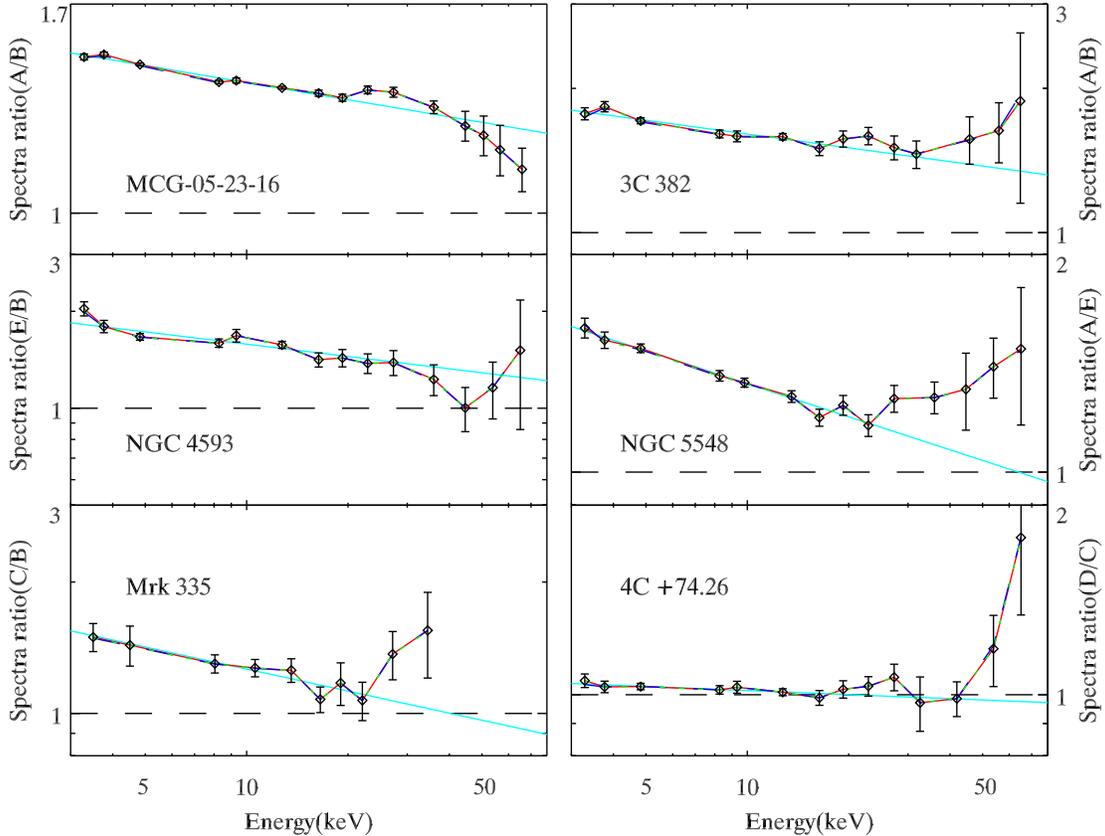}
\caption{The NuSTAR spectra ratios of the five AGNs  for which the cut-off energy variations have been claimed, plus 4C +74.26. Data points and the red dashed lines are derived from the 3rd approach (using single powerlaw),
while blue and green lines are from the 1st \& 2nd approaches. Note the blue, green and red lines almost completely overlap. 
Cyan solid lines plot best-fit powerlaw to the ratio spectra in the range of 3 -- 20 keV. 
}
\label{osufs}
\end{figure*}

We re-investigate NuSTAR spectra of these sources and examine their spectra ratios. 
For each object which have been observed more than twice by NuSTAR, we pick the two observations between which the most prominent variation of cut-off energy was claimed
(see Table.\ref{6AGNs}), to plot 
the resulted spectra ratios (of the brighter epoch to the fainter one, Fig.~\ref{osufs}) for these 5 AGNs plus 4C +74.26 (a new source with $E_c$ variation detected by this work). 
The spectra ratio plotted was obtained with the 3rd approach (using single powerlaw). 
The 1st and 2nd approaches yield practically identical results (Fig. \ref{osufs}).

The spectra ratios clearly tell the spectral slope variations: 
the ratios generally decrease at higher energies, indicating all six sources follow the well known softer-when-brighter pattern. 
We fit the ratio spectra between 3 -- 20 keV with a single power law to demonstrate the changes of the spectral slopes. 

Upward curvatures at high energies in the spectra ratios are seen in five objects  3C 382, NGC 4593, NGC 5548, Mrk 335, 4C +74.26, suggesting possible higher E$_c$ in the brighter and softer epochs (hotter-when-brighter). Meanwhile for MCG -5-23-16 a downward curvature is seen.
However, as we will show below in details, the high energy curvatures in the spectra ratio plots are inconsistent with the claimed $E_c$ ($T_e$) variation patterns in literature in three sources,
including MCG -5-23-16, 3C 382 and NGC 4593. Their $E_c$ variation patterns thus need to be revised.

\begin{figure}
\centering
\includegraphics[width=0.5\textwidth]{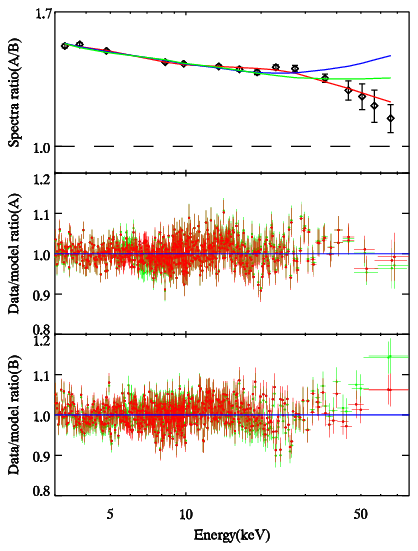}
\caption{Upper panel: the spectra ratio plot (data points, derived from the 3rd approach) of two NuSTAR exposures of MCG -5-23-16, and the ratios of best fit models to two exposures. Blue: the ratio of two best-fit models (to two observations) of \cite{Zoghbi2017}; Green: the ratio of our independent  best-fit models following \cite{Zoghbi2017}; Red: the ratio of our revised best-fit models of this work. Middle and lower panels: the data to best-fit model ratios for two NuSTAR observations (Obs. A: Obs.60001046002,
and B: Obs.60001046004). Green: data to model ratio following \cite{Zoghbi2017}, and Red: using our revised model.}
\label{mcgratio}
\end{figure}

\begin{figure*}
\centering
\includegraphics[width=0.9\textwidth]{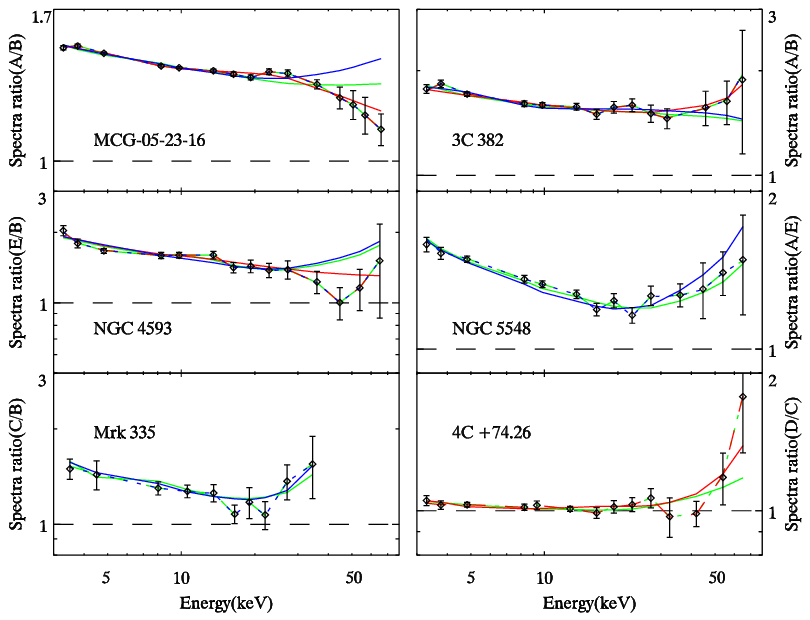}
\caption{The comparison between NuSTAR spectra ratios (data points, derived following the 3rd approach) and the ratios of best-fit models (solid lines) of AGNs for which the cut-off energy variations (or T$_e$) have been claimed using NuSTAR data. Blue solid lines plot the ratio of best-fit models reported in literature (fitting to NuSTAR spectra and data from other instruments simultaneously) which claimed the variations of E$_c$ (or T$_e$). Except for 4C +74.26, the green solid lines plot the ratios of two best-fit models derived through our independent spectral fitting using models identical to literature studies.
The weak differences between the solid blue and green lines could be attributed to slightly different source/background spectra extraction regions and/or data calibration between this study and literature.
The red solid lines plot the ratio of revised best-fit models (if needed). For 4C +74.26, the red line plots the ratio of out best-fit models. 
Note the spectra ratios obtained using the best-fit models (both the literature ones and our own) are practically identical (over-plotted with blue, green and red dashed lines, again almost totally overlap). 
}
\label{osratio}
\end{figure*}

For each source, we first adopt spectral models identical to those used in the literature to fit NuSTAR spectra (our analyses also include simultaneous  data taken by Swift or XMM-Newton, see Table.~\ref{6AGNs}). 
Note during the spectra fitting the Fe K line spectral region is included (it is only excluded in the spectra ratio plot in this work).

Below we discuss these sources one by one. 
\\

$\bullet$MCG -5-23-16:     \cite{Zoghbi2017} fitted four NuSTAR (and simultaneous Swift XRT) spectra with cutoff powerlaw plus ionized reflection {\sc xillver} and relativistic reflection component {\sc relxill}. Galactic absorption with $N_H=9\times10^{20}cm^{-2}$ and local galaxy absorption are included using the {\sc tbabs} and {\sc ztbabs} model. The model in {\sc xspec} is  {\sc tbabs$*$ztbabs$*$(cutoffpl+relxill+xillver)}. They
yielded a clear positive correlation between E$_c$ and both the hard X-ray flux and $\Gamma$ (see their Fig. 7).

In Fig. ~\ref{mcgratio} we compare the spectra ratio for NuSTAR Obs. A \& B with the ratios of best-fit models. 
Strikingly, the claimed $E_c$ variation by \cite{Zoghbi2017} (the ratio of two best-fit models, blue solid line) significantly deviates from the spectra ratio. 

We perform independent spectral fitting using identical spectral model. We fix the inclination $\theta$, disc inner radius R$_{in}$, disc emissivity index $q$ at best-fit values in \cite{Zoghbi2017} (freeing them does not significantly alter the fitting results).
Our independent spectra fitting, though yielding slightly different fitting results (the green solid line), is inconsistent with the observed spectra ratio either\footnote{Replacing {\sc RELXILL} with lamppost geometry reflection model {\sc RELXILLLP} does not alter the results.}. 
We note the slight different fitting results between us and \cite{Zoghbi2017} might be due to slightly different source/background spectra extraction regions and/or data calibration adopted. 
Residuals are clearly seen at $>$ 40 keV in the data to model plots for Obs. A and B (Fig. \ref{mcgratio})\footnote{Such deviations are also visible in Fig. 6 of \cite{Zoghbi2017}}.

We notice that the spectra ratio (Fig. ~\ref{mcgratio}) displays a bump at 20 $\sim$ 30 keV, suggesting stronger Compton reflection during the brighter epoch. However, such a bump is not seen in the ratio of best-fit models (blue and green solid lines in Fig. ~\ref{mcgratio}).
Thus revised spectral fitting is required to better interpret the spectra. 
 
The adopted disc reflection model {\sc relxill} handles the disc continuum reflection and the Fe K$\alpha$ emission line in a self-consistent manner assuming a powerlaw disc emissivity and a common ionization parameter over the while disc. This might have been over-simplified in some cases considering the corona geometry and the disc ionization could be more complicated {(such as non-axisymmetric corona illumination and inhomogeneous disc ionization).
Meanwhile, the highly ionized ultra-fast outflows detected in AGNs \citep{Tombesi2010} may also contribute to the Fe K line emission even if the outflow is out of the line of sight (with no blue-shifted absorption line detected). Note in MCG -5-23-16, a variable absorption at $\sim$ 7.7 keV was previously reported \citep{Braito2007, Tombesi2010}, suggesting the existence of such outflow. 
It is thus not surprising that such complex factors would make the self-consistent disc reflection modeling failed in some sources.
Modeling the line and continuum reflection self-consistently in such circumstances requires spectra with much higher S/N to distinguish various factors, and is beyond the scope of this work.
In such a case, fitting with disc continuum reflection and broad Fe K$\alpha$ line decoupled (hereafter the decoupled approach) could be an alternative option, which though should be used with cautious. }
Note \cite{Mantovani2016} did find that the strength of the disc continuum reflection component could be decoupled with that of the line in some sources. 
So we select to replace {\sc relxill}  with {\sc relconv$*$pexriv} plus a broad Gaussian, { where {\sc pexriv} models the continuum reflection from ionized material (\citealt{Magdziarz1995}).
} 

{ This decoupled approach yields broad Gaussian line central energy of 6.30$^{+0.04}_{-0.04}$ (6.40$^{+0.03}_{-0.02}$) keV, line width $\sigma$ of 0.38$^{+0.05}_{-0.04}$ (0.37$^{+0.04}_{-0.03}$) keV, and equivalent width of 64$^{+7}_{-5}$ (104$^{+11}_{-11}$) eV for Obs. A (B). 
We note that while the broad Fe K$\alpha$ line EW increases from 64 to 104 eV from Obs. A  to Obs. B,  the continuum reflection fraction $R$ of {\sc pexriv} drops from 0.25 to 0.14.
Meanwhile, weak line shift is seen. These facts suggest complicated and decoupled variation behaviors of the line and continuum reflection, likely due to the possibilities aforementioned.
}

{ Impressively, 
the obtained fitting results through the decoupled approach} clearly better match the spectra ratio (see Fig.~\ref{mcgratio}). The fit was also improved significantly ($\Delta \chi^2 = 27.50$ for Obs. A and $\Delta \chi^2 = 40.51$ Obs. B, for two more free parameters\footnote{{ The confidence levels of the improvements are $>$ 99.9999\%  based on F-test.}}; also see Fig.~\ref{mcgratio}). The decoupled approach yields no variation in E$_c$, but a change in the proportion of disc reflection, { the latter explains the 20 -- 30 keV bump and high energy downward curvature in the spectra ratio plot.}
 This also demonstrates that  high energy curvature in the spectra ratio plot  does not necessarily indicate $E_c$ variation, and detailed followup spectral fitting is required to confirm/quantify $E_c$ variations. 
We list the measured $\Gamma$ and E$_c$ in Table.~\ref{6AGNs} (and also for other NuSTAR observations not plotted here).  

{ We conclude that the claimed E$_c$ variation pattern in literature (hotter-when-brighter) of MCG -5-23-16 is inconsistent with the spectra ratio plot. 
Our revised spectral modeling yields statistically improved fitting results. Such results agree with the spectra ratio, but suggest no E$_c$ variation. As for E$_c$ variation studies, no variation is the null hypothesis, 
we conservatively conclude that there is no yet clear evidence of E$_c$ variation in MCG -5-23-16.
}
\\
\\

As we have shown comparing the spectra ratio with the ratio of best-fit models for two observations is useful to examine the goodness of spectral fitting, in Fig. \ref{osratio} we over-plot the spectra ratios with the ratios of the best-fit models for all six sources studies in this work.
We stress again that using the more complicated model adopted in literature and in this work yield spectra ratios practically identical to the one plotted (derived with the 3rd approach using a single powerlaw model, Fig. \ref{osratio}). This is because the spectra ratio is rather insensitive to the adopted models as we have shown with simulated spectra. 
\\
 	\begin{table*}
		\begin{center}
			\caption{ NuSTAR(NU), Swift(SW) or XMM-Newton(XMM) observations,} E$_c$ measurements, and coronal properties of the six AGN  studied in this work. \label{6AGNs}
			\begin{tabular}{ c c c c c c c} 
				\hline \rule{0pt}{4.0ex} Source & $z$ & Obs. Id. & $\Gamma$ & $\Ec^a$(keV)  & $log(F_{x})^b$(ergs $cm^{-2}$ $s^{-1}$) \\ 
				&&&&&\\
				MCG 5-23-16 & 0.009 & SW00080421008 \& NU60001046002(A) & $1.85 \pm 0.01$ & $123^{+6}_{-3}$ & -9.53 \\	
				&&&&&\\
				&  & SW00080421003 \& NU60001046004(B) & $1.75 \pm 0.01$ & $126_{-6}^{+7}$ & -9.66 \\
				&&&&&\\
				&  & SW00080421006 \& NU60001046006(C) & $1.80 \pm 0.01$ & $127^{+11}_{-8}$ & -9.59  \\
				&&&&&\\
				&   & SW00080421009 \& NU60001046008(D) & $1.75 \pm 0.01$ & $126 \pm 5$ & -9.57 \\
				&&&&&\\\hline \rule{0pt}{4.5ex}
				3C 382  & 0.058 & SW00080217001 \& NU60061286002(A) & $1.81^{+0.02}_{-0.03}$ & $(>159)365^{+546}_{-182}$& -9.79 \\
				&&&&&\\
				&&&$\tau$=$0.54^{+1.39}_{-0.24}$&$k$T$_{e}$=$121^{+62}_{-15}$&\\
				&&&&&\\
				&  & NU60001084002(B) & $1.66^{+0.01}_{-0.02}$ & $129^{+24}_{-18}$ & -10.00\\
				&&&&&\\
				&&&$\tau$=$2.28^{+0.15}_{-0.20}$&$k$T$_{e}$=$30^{+5}_{-3}$&\\	
				&&&&&\\\hline \rule{0pt}{4.5ex} 
				NGC 4593 & 0.009 & XMM0740920201 \& NU60001149002$-$\uppercase\expandafter{\romannumeral1} & $1.77\pm0.01$ & $>185(740^{+\infty}_{-472})$ & -10.11 \\
				 &&&&&\\
				&  &  XMM0740920201 \& NU60001149002$-$\uppercase\expandafter{\romannumeral2} & $1.78 \pm 0.01$ & $>50(1000^{+\infty}_{-853})$ & -10.42 \\
				&&&&&\\
				&  &  XMM0740920301 \& NU60001149004(B) & $1.67 \pm 0.01$ & $>50(334^{+307}_{-167})$ & -10.23 \\
				 &&&&&\\
				& & XMM0740920401 \& NU60001149006(C) & $1.69 \pm 0.01$ & $>50(821^{+\infty}_{-615})$ & -10.26 \\
				&&&&&\\
				&  & XMM0740920501 \& NU60001149008(D) & $1.83 \pm 0.01$ & $>447(1000^{+\infty}_{-292})$ & -10.11 \\	
				&&&&&\\
				&  & XMM0740920601 \& NU60001149010(E) & $1.84 \pm 0.01$ & $>242(942^{+\infty}_{-573})$ & -10.17 \\
				&&&&&\\\hline \rule{0pt}{4.5ex}
				NGC 5548 & 0.0172 & XMM0720110601 \& NU60002044002(A) & $1.66\pm0.03$ & $275^{+203}_{-84}$ & -9.77 \\	
				&&&&&\\
				& &  XMM0720110601 \& NU60002044003(B) & $1.70 \pm 0.03$ & $>200(429_{-175}^{+\infty})$ & -9.84 \\
				&&&&&\\
				& &  XMM072011101 \& NU60002044005(C) & $1.68 \pm 0.05$ & $148_{-52}^{+30}$ & -9.94 \\
				&&&&&\\
				& &  NU60002044006(D) & $1.83\pm0.07$ & $178_{-64}^{+181}$ & -9.93 \\
				&&&&&\\
				& & XMM0720111501 \& NU60002044008(E) & $1.47 \pm 0.03$ & $82_{-9}^{+11}$ & -9.98 \\
				 &&&&&\\\hline \rule{0pt}{4.5ex}
				Mrk 335 & 0.026 & NU60001041002(A) & $1.32\pm0.11$ & ${28^{+6}_{-10}}$ & -11.07 \\
				 &&&&&\\	
				 & &  NU60001041003(B) & $1.82_{-0.12}^{+0.15}$ & $62^{+20}_{-17}$ & -10.94 \\
				  &&&&&\\	
				 & &  NU60001041005(C) & $2.01_{-0.16}^{+0.04}$ & $>175(382^{+\infty}_{-158})$ & -10.75 \\
				&&&&&\\
				& &  NU80001020002(D) & $2.03_{-0.06}^{+0.05}$ & $85^{+25}_{-12}$ & -10.79 \\
				&&&&&\\\hline \rule{0pt}{4.5ex}
				 4C +74.26 & 0.104 & SW00080795001 \& NU60001080002(A) & $1.84 \pm 0.03$ & $156^{+74}_{-40}$ & -10.04 \\
				 &&&&&\\
				 & & SW00080795002 \& NU60001080004(B) & $1.82 \pm 0.02 $ & $172^{+35}_{-45}$  &-10.01\\
				 &&&&&\\
				 & & SW00080795003 \& NU60001080006(C) & $1.80 \pm 0.02$ & $130^{+32}_{-18}$ & -10.04 \\
				&&&&&\\
				&  &  SW00080795004 \& NU60001080008(D) & $1.86^{+0.02}_{-0.01}$ & $283_{-73}^{+174}$ & -10.00  \\	
				&&&&&\\	
				\hline  
			\end{tabular}
		\end{center}
		$^a$For non-detections of E$_c$, 90\% confidence level lower limits are given, together with the best-fit values and 1$\sigma$ confidence ranges (in parentheses).\\
		$^b$0.1 -- 200 keV flux.\\
\end{table*}

$\bullet$  3C 382:  Through fitting two NuSTAR spectra with thermal  Comptonization model {\sc compPS} \citep{Poutanen1996},  
\cite{Ballantyne2014} claimed the corona temperature $k$T$_e$ increased from 231$^{+50}_{-88}$ keV in the high flux observation (NuSTAR Obs. A)
to 330 $\pm$ 30 keV in the low flux data (NuSTAR Obs. B, see Table.~\ref{6AGNs}). 
Galactic absorption with $N_H = 6.98\times10^{20} cm^{-2}$ and a ionized warm absorber with $N_H = 1.4\times10^{21} cm^{-2}$ and $log\xi=2.5$ and a narrow FeK$\alpha$ line fixed at 6.4 keV are included in model. 
However, this model appears inconsistent with the spectra ratio at $>$ 30 keV (Fig. \ref{osufs} \& \ref{osratio}), the latter suggests a higher E$_c$ in Obs. A instead.

\begin{figure}
\centering
\includegraphics[width=0.5\textwidth]{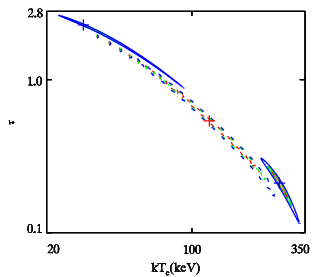}
\caption{The contour plots of $compPS$ parameter optical depth $\tau$ and corona temperature $k$T$_e$, fitting to two NuSTAR observations of 3C 382. Red, green and blue lines correspond to 1, 2 and 3$\sigma$ confidence levels, respectively. The blue crosses are the two local minimums of the low flux observation (Obs. B). The red cross and dashed contours are from the high flux observation (Obs. A).} 
\label{_3c382_contour}
\end{figure}

Through fitting the same spectra with identical model, we find significant degeneracy between optical depth $\tau$ and corona temperature $k$T$_e$ (see Fig.~\ref{_3c382_contour})\footnote{The effect of such degeneracies \citep[see also][]{Brenneman2014} should be examined while interpreting the observed $T_e$ -- $\tau$ relation in AGN samples \citep{Tortosa18}.}. 
For the low flux observation (Obs. B), 
spectral fitting yields two local minimums, one at $k$T$_e$ $\sim$ 263 keV ($\chi^2$/dof  = 644.1/673), slightly lower than but statistically consistent with that (330 $\pm$ 30) reported by \cite{Ballantyne2014},  and another at $k$T$_e$ = 30$^{+5}_{-3}$ keV ($\chi^2$/dof  = 650.3/673).  Though statistically slightly worse ($\Delta\chi^2$ = 6.2), the small $k$T$_e$ solution appears more consistent with the spectra ratio (see Fig.~\ref{osratio}), comparing with the high $k$T$_e$ one.

For the high flux observation (Obs. A), the degeneracy between $\tau$ and $k$T$_e$ is also visible (Fig.~\ref{_3c382_contour}), but only one global minimum was reached. 
The best-fit $k$T$_e$ is 121$^{+62}_{-15}$ keV, lower than but statistically consistent with that (231$^{+50}_{-88}$ keV) reported by \cite{Ballantyne2014}.

We also fit two spectra with cutoff powerlaw plus relativistic disc reflection component ({\sc relxill}), and obtain an E$_c$ of 365$^{+546}_{-182}$ keV for the high flux observation, and 129$^{+24}_{-18}$ keV for the low flux exposure (Table.~\ref{6AGNs})\footnote{\cite{Ballantyne2014} also fitted the spectra with cut-off powerlaw model and yielded E$_c$ ($214^{+147}_{-63}$ keV for low flux observation and $>190$ keV for high flux observation).}, also suggesting lower corona temperature during the low flux exposure (though statistically marginal).
Meanwhile, note that the high $k$T$_e$  solution ($k$T$_e$ $\sim$ 263 keV, while E$_c$ $\sim$ 129 keV) for the low flux observation appears incompatible  with the common approximation between $k$T$_e$ and E$_c$ ($k$T$_e$ $\approx$ E$_c/3$ for $\tau$ $\gg$ 1, and $k$T$_e$ $\approx$ E$_c/2$ for $\tau$ $\lesssim$ 1).

We conclude that for Obs. B, the low $k$T$_e$ solution is  more reasonable, and 3C 382 shows higher corona temperature (instead of lower corona temperature) at higher X-ray flux (Table.~\ref{6AGNs}).
\\

$\bullet$ NGC 4593: \cite{Ursini2016}  jointly fitted 5 pairs of simultaneous NuSTAR and XMM-Newton spectra of NGC 4593, and found E$_c$
varies from as low as 90$^{+40}_{-20}$ keV to $>$ 700 keV. 

Following \cite{Ursini2016}, we adopt a cut-off powerlaw to fit the NuSTAR and XMM spectra simultaneously, and model the reflection component with {\sc relxill} and {\sc xillver}. A Galactic absorption with $N_H = 1.88\times10^{20} cm^{-2}$ and two warm absorbers are included in model.
We yield almost identical best-fit results, but the spectra ratio plot appears inconsistent with that of the best-fit models at $>$ 30 keV,  questioning the claimed E$_c$ variation \citep{Ursini2016}.
To solve such discrepancy, we again utilize {\sc relconv$*$pexriv} plus broad Gaussian (instead of {\sc relxill}) to fit the disc reflection component. We find the revised model does yields better agreement to the spectra ratio (Fig.~\ref{osratio}),
though the spectral fitting is only slightly improved ($\Delta \chi^2 \sim 5$ for two more free parameters). 

We { conservatively} conclude that there is no evidence of significant E$_c$ variation in NGC 4593. 
Indeed, the revised spectral fittings yield no detection of E$_c$ at all in NGC 4593. 
{ The weak upward curvature in the spectra ratio plot hints possible E$_c$ variation, but can only be verified with future high quality spectra.}
\\

$\bullet$ NGC 5548: The ratio of the best-fit continuum models of \cite{Ursini2015} appears qualitatively consistent with the spectra ratio (Fig.~\ref{osratio}), indicating a clear positive E$_c$ -- $\Gamma$ correlation (see Fig. 6 of Ursini et al). Following \cite{Ursini2015}, we fit the spectra with cutoff powerlaw plus cold reflection component ({\sc pexmon}), with  two partially covering absorbers, six warm absorbers \citep{Kaastra2014b}, and soft X-ray excess included. 
Our independent spectral fitting (with simultaneous XMM spectra included) following \cite{Ursini2015} yields slightly different measurements of E$_c$ and better agrees with the spectra ratio plot. We set the initial parameters value to the measurements by \cite{Ursini2015}, but do not find a local minimum.
 The weak deviations could be attributed to slightly different data extraction and/or calibration between this work and \cite{Ursini2015}.
Our best-fit E$_c$, $\Gamma$ and X-ray fluxes are also listed in Table.~\ref{6AGNs}.
\\

$\bullet$ Mrk 335: NuSTAR observed Mrk 335 four times between 2013 and 2014. 
Combining each of the two NuSTAR observations obtained on June 13, 2013 (with net exposure time of 21 and 22 ks) with an overlapping Suzaku exposure (with a net exposure of 144 ks), \cite{Keek2016} reported E$_c$ of 31$\pm$7 and 50$\pm$12 keV respectively (see Table.~\ref{6AGNs}). 
Meanwhile lower limits of E$_c$  ($>$ 300 keV) were reported for two other NuSTAR observations obtained on June 25, 2013 (93 ks) and Sep. 20, 2014 (69 ks).

\cite{Keek2016} adopted cutoff powerlaw plus relativistic disc reflection  ({\sc relxill}) model plus a  narrow Fe K$\alpha$ to fit the spectra. A Galactic absorption with $N_H = 3.6\times10^{20} cm^{-2}$ is included. 
Follow \cite{Keek2016}, we fix black hole spin $a=0.89$ and iron abundance relative to solar $A_{\mathrm{Fe}}=3.9$. We perform our own spectral fitting to NuSTAR data alone (Table.~\ref{6AGNs}). The results are generally consistent with \cite{Keek2016}\footnote{Except for that we find a lower E$_c$ $\approx$ 84 keV for Obs. D (instead of $>$ 300 keV). 
We note for Obs. D,  \cite{Keek2016} obtained rather high $\chisq/\dof$ (635/563 = 1.13), while our fitting yields a much better one ($\chisq/\dof=$371/412).
The new best-fit E$_c$ we obtained for Obs. D (the 4th NuSTAR exposure ) is smaller (but likely statistically insignificant) than that from Obs. C, and larger (again statistically insignificant) that those from Obs. A \& B.}, and both our fittings and those from \cite{Keek2016} well match the spectra ratio (Fig. \ref{osratio}). 
Note the uncertainties to E$_c$ could have been underestimated as several model parameters have been simply fixed (black hole spin, iron abundance, inclination, etc.)

Based on our own spectral fitting (see Table.~\ref{6AGNs}), we conclude that Mrk 335 is hotter when it brightens in X-ray (see Fig. \ref{osratio}). 
\\

\begin{figure}
\centering
\includegraphics[width=0.5\textwidth]{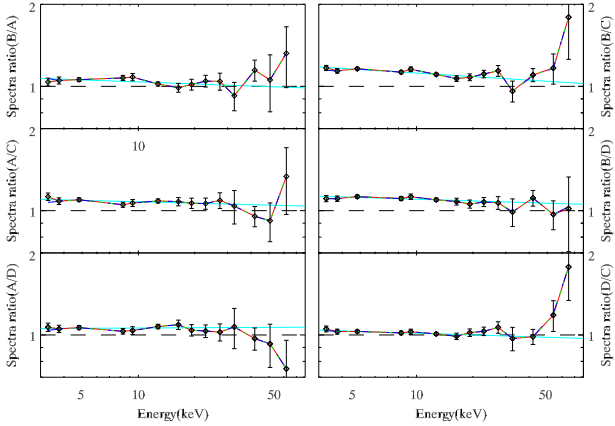}
\caption{The  spectra ratios of 4C +74.26 between each two observations. The spectra ratios derived following the 3 approaches are all plotted (blue line: the 1st approach; green line: the 2nd approach; data points and red line: the 3rd approach. The blue, green and red lines almost totally overlap).  Cyan solid line: powerlaw fit to 3 -- 20 keV of the spectra ratios.
}
\label{ratio}
\end{figure}

$\bullet$ 4C +74.26: \cite{Lohfink2017} presented spectra fitting to time-averaged NuSTAR and Swift spectra of 4C +74.26. They 
reported clear detection of high energy cut-off of 183$^{+51}_{-35}$ keV, indicating its X-ray emission is dominated by corona emission instead of jet.
In this work, the spectra from 4 NuSTAR observations were analyzed separately, and simultaneous Swift spectra (see Table 3) are included. 
In Fig.~\ref{ratio}, we show the spectra ratios of 4C +74.26 between each two observations (the brighter spectrum divide the fainter one, hereafter the same for simplicity).
Upward or downward curvatures in the plots are seen at high energies, suggesting possible E$_c$ variations between exposures.

We adopt an absorbed cutoff powerlaw and a reflection component to fit spectra. As the Fe K$\alpha$ line was found to be broad in 4C +74.26 \citep{Ballantyne2005} and is likely produced in the accretion disc, we utilize {\sc relxill} to fit the disc reflection component. Galactic absorption\footnote{https://heasarc.gsfc.nasa.gov/cgi-bin/Tools/w3nh/w3nh.pl} is fixed at $1.15 \times 10^{20} cm^{-2}$. The intrinsic cold absorption is low and poorly constrained, and is fixed at $N_{\rm H} = 0.19 \times 10^{22} cm^{-2}$ which is the averaged best-fit value from four observations. A warm absorber \citep[with parameters fixed at the values derived by][]{DiGesu2016} are included.
For {\sc relxill}, we fix a = 0 , r$_{in}$ = 6 $R_g$, r$_{out}$ = 400 $R_g$, and we leave emissivity profile index, $\xi$ and the iron abundance $A_{Fe}$ free to vary, and the cosine of inclination angle was fixed to $50^{\circ}$. 
The best-fitting parameters are listed in
Table.~\ref{6AGNs}.

\begin{figure}
\centering
\includegraphics[width=0.5\textwidth]{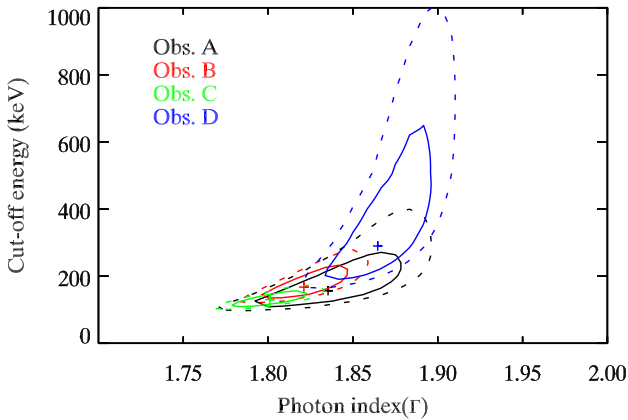}
\caption{The contours are the primary continuum cut-off energy versus photon index for each observation.  Solid and dashed lines correspond to 1, 2$\sigma$ confidence levels, respectively. The cross are the best-fit parameters in Table.~\ref{6AGNs}.}
\label{contour}
\end{figure}

The contour plots of E$_c$ versus photon index $\Gamma$ are shown in Fig.~\ref{contour}.  
The variations of the best-fit E$_c$ between observations are also consistent with what the spectra ratio plots have illustrated, that Obs. B \& D have higher E$_c$ comparing with Obs. A \& C. 
Note the upward curvature in the spectra ratio in 4C +74.26 (at $>$ 40 keV, Fig.~\ref{osratio}) appears sharper than implied by our best-fit cut-off powerlaw model.
This is likely because the corona emission produces high energy curvature sharper than a simple exponential cut-off \citep[e.g.][]{Zdziarski2003}.
Detailed analyses with Comptonization model fitting to 4C +74.26 (and other sources) are left to a future work.

\section{Discussion}\label{sect:discussion}
Studying the variations of  the X-ray high energy cutoff  of AGNs is essential to probe the corona physics. 
In this work,
we show that the  spectra ratio of two NuSTAR observations is uniquely useful in such studies. 
The spectra ratio can be utilized to easily select candidate AGNs with possible $E_c$ variations for followup detailed spectral fitting.
It is also helpful to examine the goodness of the spectral fitting and revise the model selection/settings if needed,
as the ratio of two best-fit models should agree with the spectra ratio.
This is particularly important as finding the best models which can properly interpret the spectra and the spectral variations is often not straightforward. 
In this work we collect from literature five AGNs for which $E_c$ (or corona electron temperature T$_e$) variation was claimed with NuSTAR observations(often jointly with XMM-Newton, Swift or Suzaku exposures), and re-examine their $E_c$ variations assisted with the spectra ratio technique. 
Strikingly, the E$_c$ variation patterns revealed with the spectra ratio plots contradict the literature studies for 3 of them.
Improved spectral fitting results are presented, and their $E_c$ variation patterns reported in literature are revised. 
We also report a new source (4C +74.26) with $E_c$ variations.

Bright Seyfert galaxies generally show softer X-ray spectra when they brighten in X-ray \citep{Markowitz2003b,Sobolewska2009,Ballantyne2014,Soldi2014,Fuerst2016,Ursini2016}. This trend also holds for all the six sources studies in this work, suggesting the spectral variations might
have been driven by a  common process.
The underlying physics is however unclear. 
A qualitative interpretation is that during a higher flux epoch, the corona is more effectively cooled by more seed photons.
In this scenario, a colder corona (thus higher $\Gamma$ and lower E$_c$) is expected at higher luminosity.  

However, in 4 out the 6 AGNs studies in this work (3C 382, 4C +74.26, Mrk 335 and NGC 5548) we find contrarily larger E$_c$ (or $k$T$_e$) when the spectra are brighter and softer, demanding a different interpretation.
Meanwhile we show that the rest two sources (MCG -5-23-16 and NGC 4593) show no evidence of significant E$_c$ variations. 

One possibility is geometry changes of the corona \citep[e.g.][]{Keek2016}. 
By modeling the profiles of the relativistically blurred disc reflection, studies suggested the corona of Mrk 335 has experienced geometry changes, i.e,transition from a compact source during low flux epochs to vertically extended jet-like during bright flares \citep{Wilkins2015a,Wilkins2015}. Such vertically outflowing of the corona might be common in AGNs \citep{Liu2014}. 
Interestingly, \cite{Keek2016} claimed E$_c$ of Mrk 335 is higher during high-flux epochs (through jointly fitting Suzaku \& NuSTAR spectra). They argued that the corona opacity during  high-flux epochs must be smaller, to produce steeper X-ray spectra. They suggested that at lower X-ray fluxes, the corona is compact and optically thick, located close to the inner most disc, whereas at higher accretion rates, the corona is likely optically thin and extended further way from the disc surface.
Note in this diagram, the E$_c$ variation may be contributed to stronger  general relativistic effect during low flux epochs \citep{Miniutti2004a,Tamborra18}, hotter corona during the flares, or Doppler blueshift due to corona outflowing during the flares \citep[e.g.][]{Liu2014}.

Such interpretation proposed by \cite{Keek2016} for Mrk 335 may work well for 3C 382, 4C +74.26 and NGC 5548.
Unfortunately, as a significant contribution from relativistic disc reflection is absent in NGC 5548 \citep[e.g.][]{Brenneman2012} and 3C 382 \citep{Ballantyne2014}, we are unable to testify whether corona geometry changes happened in them. For 4C +74.26, no significant variation in the ionized reflection component was detected \citep{Lohfink2017}, likely due to the limited spectra S/N. Note the equivalent width (EW) of its mildly broadened Fe K line  is slightly lower (though statistically insignificant) during Obs. B \& D ($52 \pm 13$ eV, when E$_c$ is higher) than that in Obs. A \& C ($62 \pm 10$ eV), consistent with the corona geometry change diagram. 
Meanwhile, \cite{Zoghbi2017} found the shape of the reflection spectrum in MCG -5-23-16 does not change in the two years spanned by NuSTAR observations, suggesting a stable corona geometry. 
 
 So there are two mechanisms which might be at work behind E$_c$ variations. The first is  Compton cooling which leads to a cooler corona and softer spectra during higher flux epochs.
 The second is vertically outflowing coronal flares which could convert a compact and optically thick corona to a vertically extended, outflowing and optically thin one,  thus softer spectra but higher E$_c$ are expected  at higher fluxes. 
If two mechanisms are both at work, we might expect softer spectra but no clear variation in E$_c$ at higher fluxes in many AGNs.

\cite{Fabian2015} have shown that the corona temperatures in AGNs are close to the boundary of the region in the compactness -- temperature diagram which is forbidden due to runway pair production, suggesting pair production is an essential ingredient in AGN corona.
In pair-dominated plasmas, $\Ec$ can positively correlate \citep{Ghisellini1994} with the observed photon index $\Gamma$ for electron temperatures $T_{e} < m_{e}c^{2}$ (where $m_{e}$ is the electron mass and $c$ is the speed of light), or remains constant for different values of $\Gamma$ \citep{Zdziarski2002}.

\begin{figure}
\centering
\includegraphics[width=0.45\textwidth]{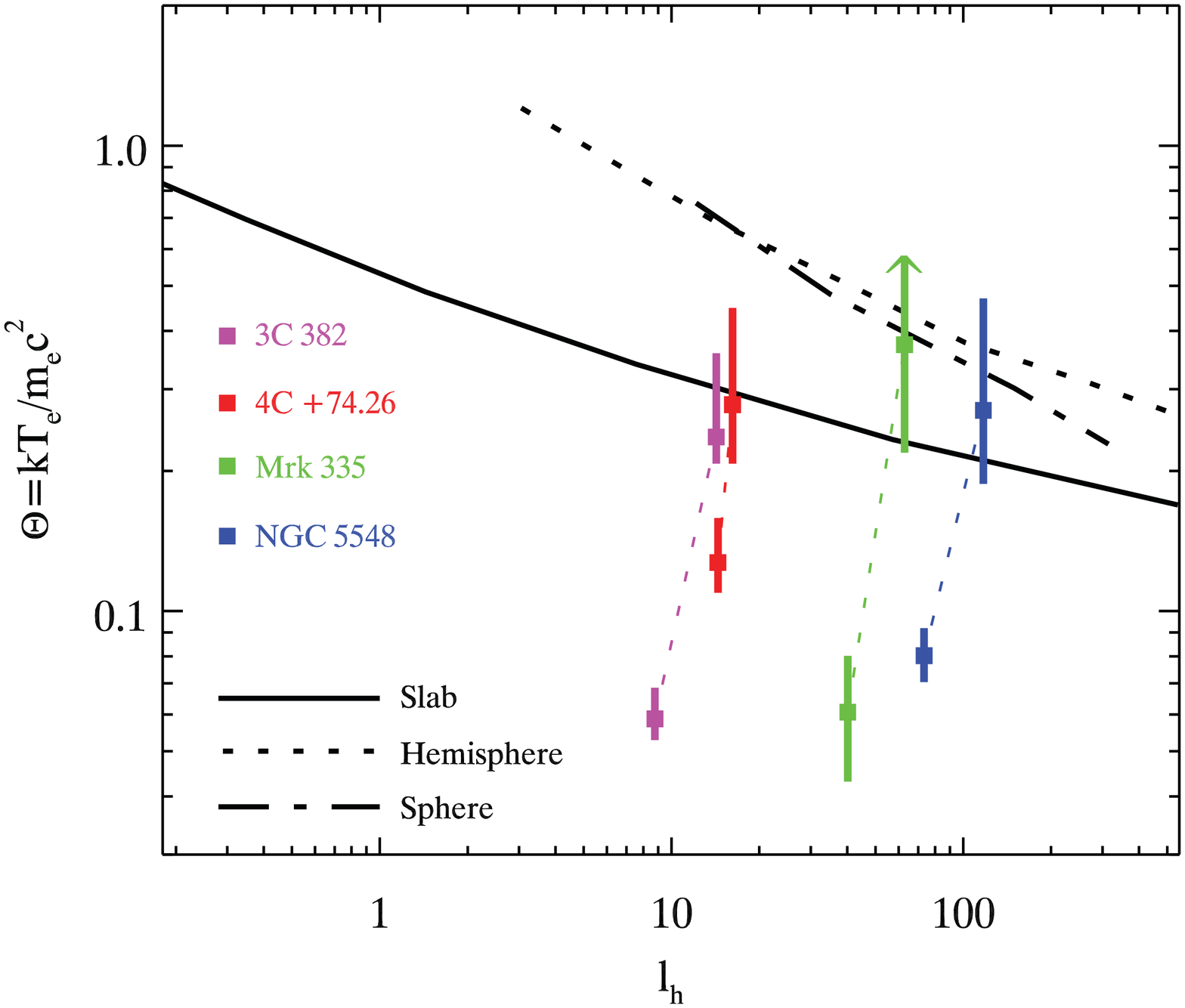}
\caption{3C 382, 4C +74.27, Mrk 335, and NGC 5548 in the $l$ -- $\Theta$ plot. For simplicity, only two NuSTAR observations (when more than two are available) between which the E$_c$ variation is the most prominent, are plotted.
The maximum temperature (in units of $m_{e}c^{2}$) that can reached by a plasma dominated by runaway pair production for three geometries \citep{Stern1995} are over-plotted as lines.}
\label{T_l}
\end{figure}

In this work, we find higher $\Ec$ (or $k$T$_e$) at larger $\Gamma$ for 3C 382, 4C +74.26, Mrk 335 and NGC 5548.
To further explore the physical nature of their spectral variations, we plot them on the compactness -- temperature diagram in Fig.~\ref{T_l}.
Pair limits from the modeling of \cite{Stern1995} for three different geometries of the corona are over-plotted. 
The compactness $l=4\pi(m_{p}/m_{e})(r_{g}/r)(L/L_{edd})$, which measures the luminosity to source size ratio, was estimated following 
a similar procedure to \cite{Fabian2015}.  For 4C +74.26, we adopt a SMBH mass of 10$^{9.45}$ $M_{\odot}$ \citep{Winter2010} and a corona size of 10 $r_g$ to calculate the compactness.
For Mrk 335, NGC 5548 and 4C +74.26, following \cite{Fabian2015} and \cite{Petrucci2001a} we estimate the electron temperature as $kT_{e} \approx E_{c}/2$,\footnote{We note that $kT_{e}$ derived through detailed Comptonization model fitting could be a better option. However, considering the strong degeneracies between Comptonization model parameters, we will leave this to a future systematic work.} and $0.1-200$ keV model luminosity is adopted to calculate $l$.
We note the coronae of 3C 382 and 4C +74.26, Mrk 335 and NGC 5548 do appear pair dominated during the brighter epochs when larger E$_c$ were detected, suggesting that pair production may play a
non-negligible role during the spectral variations. We note that in Fig. \ref{T_l} a constant corona size was adopted. Considering the corona during the higher E$_c$ epochs could be more extended, as discussed above, we need shift their upper $\Theta$ data points in Fig. \ref{T_l} to the left. However, a shift as large as one magnitude  does not significantly change the conclusion that their corona are pair dominated during the higher E$_c$ epochs.

\section*{Acknowledgment}
We especially thank the anonymous referee for his/her helpful comments and suggestions that have significantly improved the paper.
This work is supported by National Basic Research Program of China (973 program, grant No. 2015CB857005) and National Science Foundation of China (grants No. 11421303).
J.X.W. thanks support from Chinese Top-notch Young Talents Program, and  CAS Frontier Science Key Research Program QYCDJ-SSW-SLH006.
\end{CJK*}

\bibliography{ref}
\end{document}